\documentclass[twocolumn,showpacs,aps,prd]{revtex4-1}

\usepackage{graphicx}
\usepackage{dcolumn}
\usepackage{amsmath}
\usepackage{epsfig}
\usepackage{subfigure}
\usepackage{epstopdf}
\usepackage{multirow}
\usepackage{mathrsfs}
\usepackage[bookmarksnumbered, pdfstartview=FitH,colorlinks,urlcolor=blue, citecolor=blue,linkcolor=blue,] {hyperref}

\usepackage{lineno}
\usepackage{enumerate}
\usepackage{float}
\usepackage{color}

\begin{document}

\title{\bf \boldmath
Search for the hyperon semileptonic decay $\Xi^{-} \rightarrow \Xi^{0} e^{-} \bar{\nu_{e}}$
}

\author{
\begin{small}
\begin{center}
M.~Ablikim$^{1}$, M.~N.~Achasov$^{10,b}$, P.~Adlarson$^{68}$, S. ~Ahmed$^{14}$, M.~Albrecht$^{4}$, R.~Aliberti$^{28}$, A.~Amoroso$^{67A,67C}$, M.~R.~An$^{32}$, Q.~An$^{64,50}$, X.~H.~Bai$^{58}$
% ...
\\
\vspace{0.2cm}
(BESIII Collaboration)\\
\vspace{0.2cm} {\it
$^{1}$ Institute of High Energy Physics, Beijing 100049, People's Republic of China\\
% ...
$^{4}$ Bochum Ruhr-University, D-44780 Bochum, Germany\\
% ...
$^{10}$ G.I. Budker Institute of Nuclear Physics SB RAS (BINP), Novosibirsk 630090, Russia\\
% ...
$^{14}$ Helmholtz Institute Mainz, Staudinger Weg 18, D-55099 Mainz, Germany\\
% ...
$^{28}$ Johannes Gutenberg University of Mainz, Johann-Joachim-Becher-Weg 45, D-55099 Mainz, Germany\\
% ...
$^{32}$ Liaoning Normal University, Dalian 116029, People's Republic of China\\
% ...
$^{50}$ State Key Laboratory of Particle Detection and Electronics, Beijing 100049, Hefei 230026, People's Republic of China\\
% ...
$^{58}$ University of Jinan, Jinan 250022, People's Republic of China\\
% ...
$^{64}$ University of Science and Technology of China, Hefei 230026, People's Republic of China\\
% ...
$^{67}$ University of Turin and INFN, (A)University of Turin, I-10125, Turin, Italy; (B)University of Eastern Piedmont, I-15121, Alessandria, Italy; (C)INFN, I-10125, Turin, Italy\\
$^{68}$ Uppsala University, Box 516, SE-75120 Uppsala, Sweden\\
% ...
$^{b}$ Also at the Novosibirsk State University, Novosibirsk, 630090, Russia\\
% ...
}
\end{center}
\vspace{0.4cm}
\end{small}
}

\begin{abstract}
Using $(10.087\pm0.044)\times10^{9}$ $J/\psi$ events collected by the BESIII detector at the BEPCII collider, we search for the hyperon semileptonic decay $\Xi^{-} \rightarrow \Xi^{0} e^{-} \bar{\nu}_{e}$. No significant signal is observed and the upper limit on the branching fraction $\mathcal B(\Xi^{-} \rightarrow \Xi^{0} e^{-} \bar{\nu}_{e})$ is set to be $2.59\times10^{-4}$ at 90\% confidence level. This result is one order of magnitude more strict than the previous best limit.
\end{abstract}

\maketitle

\section{\boldmath INTRODUCTION}

Studies of hyperon semileptonic decays provide important information on the interplay between weak interactions and hadronic structures formed through strong interactions.
Semileptonic decays of baryons provide richer information than those of mesons due to the presence of three valence quarks rather than a quark$-$antiquark pair~\cite{2007BatleyPLB645}.
In addition, it has previously been shown that flavor SU(3) symmetry is manifestly broken in hyperon semileptonic decays~\cite{2015YangPRC92}.
Therefore, with more complete information on hyperon semileptonic decays, the patterns of flavor SU(3) symmetry breaking could be further revealed in nature~\cite{2015YangPRC92, 2013PhamPRD87}.

Since the branching fractions of hyperon semileptonic decays are on the order of $10^{-4}$ or smaller~\cite{2020PdgPTEP2020}, studies of hyperon semileptonic decays are still an experimental challenge. Except for the measurements performed by the KTeV and NA48/1 Collaborations of $\Xi^{0} \rightarrow \Sigma^{+}\ell\bar{v}_{\ell}$ ($\ell=e,\mu$) decays~\cite{2005AbouzaidPRL95, 2013BatleyPLB720}, most hyperon semileptonic results are more than 30 years old~\cite{2020PdgPTEP2020, 2015ChangPRL114}. There is thus much room for improvement on the experimental side~\cite{2017LiFP12}.

The hyperon semileptonic decay $\Xi^{-} \rightarrow \Xi^{0} e^{-} \bar{\nu}_{e}$ has not yet been observed~\cite{2020PdgPTEP2020}. Previously, an experiment at Brookhaven National Laboratory (BNL)~\cite{1974YehPRD10} set an upper limit of $2.3 \times 10^{-3}$ on the branching fraction
$\mathcal B(\Xi^{-} \rightarrow \Xi^{0} e^{-} \bar{\nu}_{e})$ at 90$\%$ confidence level (C.L.) based on 8150 $\Xi^{-}$ events.
The BESIII experiment~\cite{Ablikim:2009aa} has recently collected $(10.087\pm0.044)\times10^{9}$ $J/\psi$ events, which is the world's largest data sample of $J/\psi$ mesons produced in $e^{+}e^{-}$ annihilation. The total number of $J/\psi$ events collected in the years of 2009, 2012, 2018 and 2019 is determined using inclusive $J/\psi$ decays with the method described in Ref.~\cite{njpsi2017}. Within this sample, a large number of $\Xi^{-}$ events ($\sim10^{6}$~\cite{2017LiFP12}) are produced via the decay mode $J/\psi \rightarrow \Xi^{-}\bar{\Xi}^{+}$ and the expected sensitivity on the branching fraction will be on the order of $10^{-4}$, thereby providing a good opportunity to study this hyperon semileptonic decay. The theoretical prediction of the branching fraction of $\Xi^{-} \rightarrow \Xi^{0} e^{-} \bar{\nu}_{e}$  is on the order of $10^{-10}$~\cite{2019YangPRD92}, and a theoretical calculation~\cite{2015YangPRC92} shows that the effect of flavor SU(3) symmetry breaking is particularly evident in the $\Xi^{-} \rightarrow \Xi^{0} e^{-} \bar{\nu}_{e}$ decay.

In this paper, we report a search for the rare hyperon semileptonic decay $\Xi^{-} \rightarrow \Xi^{0} e^{-} \bar{\nu}_{e}$ by
analyzing $10^{6}$ $J/\psi \rightarrow \Xi^{-}\bar{\Xi}^{+}$ events collected at a center-of-mass (CM) energy $\sqrt{s} = 3.097$~${\rm GeV}/c^{2}$ with the BESIII detector.

\section{\boldmath{DETECTOR AND MONTE CARLO SIMULATION}}
The BESIII detector records symmetric $e^+e^-$ collisions provided by the BEPCII~\cite{Yu:IPAC2016-TUYA01} storage ring, which operates with a peak luminosity of $1\times10^{33}$~cm$^{-2}$s$^{-1}$ in the CM energy range from 2.0 to 4.95~${\rm GeV}/c^{2}$. BESIII has collected large data samples in this energy region~\cite{Ablikim:2019hff}. The cylindrical core of the BESIII detector covers 93\% of the full solid angle and consists of a helium-based multilayer drift chamber~(MDC), a plastic scintillator time-of-flight system~(TOF), and a CsI(Tl) electromagnetic calorimeter~(EMC), which are all enclosed in a superconducting solenoidal magnet
providing a 1.0~T (0.9~T in 2012) magnetic field. The solenoid is supported by an octagonal flux-return yoke with resistive plate counter muon identification modules interleaved with steel. The charged-particle momentum resolution at $1~{\rm GeV}/c$ is $0.5\%$, and the $dE/dx$ resolution is $6\%$ for electrons from Bhabha scattering. The EMC measures photon energies with a resolution of $2.5\%$ ($5\%$) at $1$~${\rm GeV}$ in the barrel (end cap) region. The time resolution in the TOF barrel region is 68~ps, while that in the end cap region is 110~ps. The end cap TOF system was upgraded in 2015 using multi-gap resistive plate chamber technology, providing a time resolution of 60~ps~\cite{etof}.

Simulated data samples produced with a {\sc geant4}-based~\cite{geant4} Monte Carlo (MC) package, which includes the geometric description of the BESIII detector and the detector response, are used to determine detection efficiencies and to estimate backgrounds. The simulation models the beam energy spread and initial state radiation in the $e^+e^-$ annihilations with the generator {\sc kkmc}~\cite{ref:kkmc}.
The inclusive MC sample includes both the production of the $J/\psi$ resonance and the continuum processes incorporated in {\sc
kkmc}~\cite{ref:kkmc}. The known decay modes are modeled with {\sc evtgen}~\cite{ref:evtgen} using branching fractions taken from the Particle Data Group (PDG)~\cite{2020PdgPTEP2020}, and the remaining unknown charmonium decays are modeled with {\sc lundcharm}~\cite{ref:lundcharm}. Final state radiation~(FSR) from charged final state particles is incorporated using {\sc photos}~\cite{photos}. To determine the detection efficiency, a signal MC sample with the decay chain of $J/\psi \rightarrow \Xi^{-}\bar{\Xi}^{+}$, $\bar{\Xi}^{+} \rightarrow\bar{\Lambda}\pi^{+}$, $\Xi^{-} \rightarrow \Xi^{0}(\rightarrow \Lambda\pi^{0}) e^{-} \bar{\nu}_e $ is produced, where the $\bar{\Xi}^{+} \rightarrow\bar{\Lambda}\pi^{+}$ decay is generated with the measured parameter in Ref.~\cite{2021sigPara} by BESIII experiment, and the $\Xi^{-} \rightarrow \Xi^{0} e^{-} \bar{\nu}_e$ decay is generated with a uniform distribution over the phase space.

\section{\boldmath{EVENT SELECTION AND DATA ANALYSIS}}
\label{evt_sel}

\subsection{\boldmath{Analysis method}}
\label{method}

The $\Xi^{-}$ sample is obtained via the decay mode $J/\psi \rightarrow \Xi^{-}\bar{\Xi}^{+}$. To determine the absolute branching fraction of $\Xi^{-} \rightarrow \Xi^{0} e^{-} \bar{\nu}_e$ and reduce systematic uncertainties, a tagging technique is adopted, which was first introduced by the MARK-III collaboration~\cite{1986DTPRL56}. First, one $\bar{\Xi}^{+}$ hyperon is fully reconstructed via the hadronic decay mode $\bar{\Xi}^{+} \rightarrow \bar{\Lambda}\pi^{+}$ with a large branching fraction (99.887$\pm$0.035)\%~\cite{2020PdgPTEP2020}, and then the signal decay $\Xi^{-}\rightarrow \Xi^{0} e^{-} \bar{\nu}_e$ with $\Xi^{0} \rightarrow \Lambda \pi^{0}$ is searched for in the recoiling side of the tagged $\bar{\Xi}^{+}$. The tagged $\bar{\Xi}^{+}$ events are referred to as ``single tag"~(ST) events, while the events in which the $\Xi^{-}$ semileptonic decay of interest and the ST $\bar{\Xi}^{+}$ are simultaneously found are referred to as ``double tag"~(DT) events. The absolute branching fraction is calculated by
\begin{equation}
\label{tag_method}
  \mathcal{B}(\Xi^{-} \rightarrow \Xi^{0} e^{-} \bar{\nu}_{e}) = \frac{N_{\textrm{DT}}^{\textrm{obs}}\cdot\epsilon_{\textrm{ST}}}{N_{\textrm{ST}}^{\textrm{obs}}\cdot\epsilon_{\textrm{DT}}\cdot\mathcal{B}(\Xi^{0} \rightarrow \Lambda \pi^{0}\rightarrow p\pi^{-}\gamma\gamma)},
\end{equation}
where $N_{\textrm{ST}}^{\textrm{obs}}$ ($N_{\textrm{DT}}^{\textrm{obs}}$) is the ST (DT) yield, $\epsilon_{\textrm{ST}}$ ($\epsilon_{\textrm{DT}}$) is the ST (DT) efficiency, not including the branching fractions of the subsequent decays of the $\bar{\Xi}^{+}$ ($\Xi^{0}$), and $\mathcal{B}(\Xi^{0} \rightarrow \Lambda \pi^{0}\rightarrow p\pi^{-}\gamma\gamma)$ is the branching fraction of the $\Xi^{0} \rightarrow \Lambda \pi^{0}\rightarrow p\pi^{-}\gamma\gamma$ decay.

\subsection{\boldmath{ST event selection}}
\label{ST_sel}

Charged tracks detected in the MDC are required to have a polar angle ($\theta$) satisfying $|\!\cos\theta|<0.93$, where $\theta$ is defined with respect to the beam direction. Particle identification~(PID) for charged tracks combines measurements of the $dE/dx$ in the MDC and the flight time in the TOF. The PID confidence levels are calculated for the proton (${\rm CL}_{p}$), pion (${\rm CL}_{\pi}$), and kaon (${\rm CL}_{K}$) hypotheses. The proton (pion) candidate is chosen so that the
proton (pion) hypothesis has the highest PID confidence level among these three hypotheses.

To reconstruct the $\bar{\Lambda}$ and $\bar{\Xi}^{+}$ candidates, a secondary vertex fit~\cite{2009XuCPC33} is applied to the $\bar{p}\pi^{+}$ combination and the $\bar{\Lambda}\pi^{+}$ combination, respectively, given that $\bar{\Lambda}$ and $\bar{\Xi}^{+}$ are long-lived particles. The secondary vertex fit is performed using the parameters of the production vertex, decay vertex,
and the $\bar{\Lambda}$ ($\bar{\Xi}^{+}$) flight direction.
To suppress background from non-$\bar{\Lambda}$ (non-$\bar{\Xi}^{+}$) processes, the decay length~\cite{2009XuCPC33} of the $\bar{\Lambda}$ ($\bar{\Xi}^{+}$) is required to be larger than zero, where the decay length is the distance from the production vertex to the decay vertex, and negative decay lengths can be caused by the detector resolution.
The invariant masses of the $\bar{p}\pi^{+}$ and $\bar{\Lambda}\pi^{+}$ combinations are required to satisfy $|M_{\bar{p}\pi^{+}}-M_{\bar{\Lambda}}|<0.005$~${\rm GeV}/c^{2}$ and $|M_{\bar{\Lambda}\pi^{+}}-M_{\bar{\Xi}^{+}}|<0.005$~${\rm GeV}/c^{2}$, respectively, where $M_{\bar{\Lambda}}$ ($M_{\bar{\Xi}^{+}}$) is the known mass of the $\bar{\Lambda}$ ($\bar{\Xi}^{+}$)~\cite{2020PdgPTEP2020}.
The recoiling mass against the reconstructed $\bar{\Xi}^{+}$ candidate is defined as $M_{\bar{\Lambda}\pi^{+}}^{\textrm{recoil}} \equiv \sqrt{(E_{\textrm{CM}}-E_{\bar{\Lambda}\pi^{+}})^{2}-|\vec{p}_{\bar{\Lambda}\pi^{+}}|^{2}}$, where $E_{\textrm{CM}}$ is the CM energy, and $E_{\bar{\Lambda}\pi^{+}}$ and $\vec{p}_{\bar{\Lambda}\pi^{+}}$ are the energy and momentum of the selected $\bar{\Lambda}\pi^{+}$ candidate in the CM system. To further suppress backgrounds, the recoiling mass is required to satisfy $1.290$~${\rm GeV}/c^{2}$$<M_{\bar{\Lambda}\pi^{+}}^{\textrm{recoil}}<1.342$~${\rm GeV}/c^{2}$, which corresponds to a three standard-deviation range of the $M_{\bar{\Lambda}\pi^{+}}^{\textrm{recoil}}$ distribution.

\begin{figure}[tb]
\centering
   \includegraphics[width=0.45\textwidth]{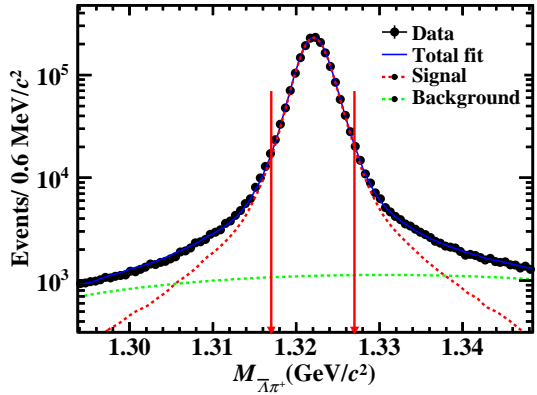}
  \caption{Fit to the invariant mass distribution of the $\bar{\Lambda}\pi^{+}$ candidates, where the black points with error bars are data, and the red dashed line and green dashed line are the signal shape and background shape, respectively. The blue solid line shows the total fit and the pair of red arrows indicates the ST signal region.}
\label{fig:ST_mxi}
\end{figure}

The $\bar{\Lambda}\pi^{+}$ candidate is required to fall in the mass window [1.317, 1.327]~${\rm GeV}/c^{2}$, which is defined as the ST signal region. This mass window corresponds to around a three standard-deviation range of the $M_{\bar{\Lambda}\pi^{+}}$ distribution. To extract the ST yield, we perform a binned maximum likelihood fit to the data distribution of the invariant mass of $\bar{\Lambda}\pi^{+}$, as shown in Fig.~\ref{fig:ST_mxi}. In the fit, the signal shape is modeled by the MC-simulated shape convolved with a Gaussian function to account for the resolution difference between data and simulation samples. By analyzing the inclusive MC samples with the help of a generic event type analysis tool, TopoAna~\cite{2021ZhouCPC258}, there is no peaking background, and the background shape is described with a second-order Chebychev function.
The ST yield extracted from the fit is $1,780,070\pm 1366$. The ST efficiency is $(24.64\pm0.02)\%$, where the uncertainty is statistical only.

\subsection{\boldmath{DT event selection}}
\label{signal_sel}

To identify the semileptonic decay $\Xi^{-}\rightarrow \Xi^{0} e^{-} \bar{\nu}_e$, we search for the $\Xi^{0}$ in the recoiling side of the ST $\bar{\Xi}^{+}$ candidates.
The $\Xi^{0}$ baryon is reconstructed via $\Xi^{0} \rightarrow \Lambda \pi^{0}$, $\Lambda \rightarrow p \pi^{-}$, $\pi^{0} \rightarrow \gamma\gamma$. Due to the very limited phase space ($M_{\Xi^{-}}-M_{\Xi^{0}}\simeq 6.85$~${\rm MeV}/c^{2}$)
and the small momentum of the $\Xi^{-}$, electrons have too small momenta to be reconstructed in the detector. To suppress backgrounds, the total number of charged tracks, including the charged tracks on the ST side, is required to be 5. Photon candidates are identified using showers in the EMC. The deposited energy of each shower must be more than 25~${\rm MeV}/c^{2}$ in the barrel region ($|\!\cos\theta|<0.80$) and more than 50~${\rm MeV}/c^{2}$ in the end cap region ($0.86<|\!\cos\theta|<0.92$). To exclude showers that originate from charged tracks, the angle between the position of each shower in the EMC and the closest extrapolated charged track must be greater than $10^{\circ}$. To suppress electronic noise and showers unrelated to the event, the difference between the EMC time and the event start time is required to be within [0, 700] ns. The $\pi^{0}$ candidates are reconstructed with a pair of photons. Due to the poor resolution in the end cap regions of the EMC, $\pi^{0}$ candidates with two daughter photons found in the end caps are rejected. The invariant mass of the two photons is required to be within $(0.115, 0.150)$~${\rm GeV}/c^{2}$. A mass-constrained kinematic fit is performed by constraining the invariant mass of $\gamma\gamma$ to the known $\pi^{0}$ mass~\cite{2020PdgPTEP2020}.

To reconstruct $\Lambda$ candidates, a vertex fit is applied to $p\pi^{-}$ combinations, and the one closest to the known $\Lambda$ mass ($M_{\Lambda}$)~\cite{2020PdgPTEP2020} is retained. The invariant mass of $p\pi^{-}$ is required to satisfy $|M_{p\pi^{-}}-M_{\Lambda}|<0.005$~${\rm GeV}/c^{2}$. The $\Xi^{0}$ is reconstructed with the $\Lambda\pi^{0}$ combinations, and the one closest to the known $\Xi^{0}$ mass ($M_{\Xi^{0}}$)~\cite{2020PdgPTEP2020} is retained for further analysis. The invariant mass of $\Lambda\pi^{0}$ is required to satisfy $|M_{\Lambda\pi^{0}}-M_{\Xi^{0}}|<0.0145$~${\rm GeV}/c^{2}$. To further suppress backgrounds, the momentum of the reconstructed $\Xi^{0}$ is required to be within $(0.79, 0.84)$~${\rm GeV}/c$, which is optimized using the Punzi figure of merit with a formula of $\epsilon/(1.5+\sqrt{B})$~\cite{2003Punzi}, where $\epsilon$ denotes the efficiency of the signal and $B$ is the number of background events.

\begin{figure}[tb]

  \includegraphics[width=0.45\textwidth]{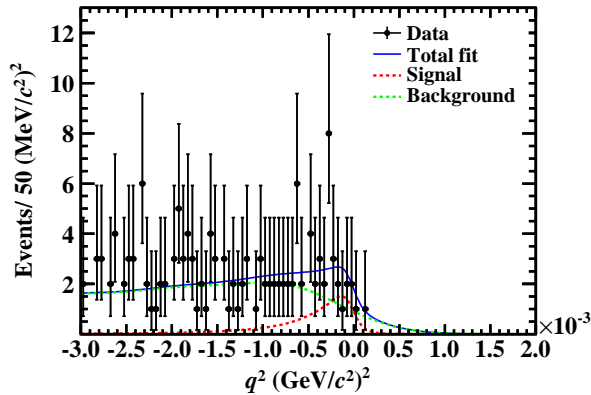}
  \caption{Fit to the $q^{2}$ distribution of the signal candidates, where the black points with error bars are data, and
the red dashed line and green dashed line are the signal shape and background shape, respectively. The blue solid line shows the total fit.}
\label{fig:DT_q2_fit}
\end{figure}

To extract the DT yield, the invariant mass squared of the lepton-neutrino system, $q^{2}$, is defined as
$q^{2} \equiv (E_{\textrm{CM}}-E_{\bar{\Xi}^{+}}-E_{\Xi^{0}})^{2}-(\vec{p}_{\textrm{CM}}-\vec{p}_{\bar{\Xi}^{+}}-\vec{p}_{\Xi^{0}})^{2}$,
where $E_{\textrm{CM}}$, $E_{\bar{\Xi}^{+}}$, and $E_{\Xi^{0}}$ are the energies of the CM, $\bar{\Xi}^{+}$, and $\Xi^{0}$, respectively, and $\vec{p}_{\textrm{CM}}$, $\vec{p}_{\bar{\Xi}^{+}}$, and $\vec{p}_{\Xi^{0}}$
are the momenta of the CM, $\bar{\Xi}^{+}$, and $\Xi^{0}$, respectively. After all the above selection criteria are applied, the DT efficiency obtained from the signal MC sample is $(2.89\pm0.01)\%$, where the uncertainty is statistical only. By analyzing the inclusive MC samples with TopoAna, the dominant background is found to be the process
$J/\psi \rightarrow \Xi^{-}\bar{\Xi}^{+}$ with $\Xi^{-} \rightarrow \Lambda\pi^{-}$ and $\bar{\Xi}^{+} \rightarrow \bar{\Lambda}\pi^{+}$. An exclusive MC sample is generated to study this dominant background and extract the background shape. To obtain the DT yield, an unbinned maximum likelihood fit is performed to the data distribution of $q^{2}$, as shown in Fig.~\ref{fig:DT_q2_fit}, where the signal shape is modeled by the MC-simulated shape, and the background shape is modeled by the exclusive MC-simulated shape. No obvious signal is observed.

\section{\boldmath SYSTEMATIC UNCERTAINTIES}
\label{Sys_err}
The systematic uncertainties in the measurement of $\mathcal B(\Xi^{-} \rightarrow \Xi^{0} e^{-} \bar{\nu}_{e})$ mainly originate from the tracking efficiency, PID efficiency, $\Lambda$ vertex fit, photon detection, $\pi^{0}$ reconstruction, tag efficiency bias, mass window of $\Lambda/\Xi^{0}$, requirement on $p_{\Xi^{0}}$, fitting range of $q^{2}$, and cited branching fractions. These systematic uncertainties are summarized in Table~\ref{tab:sum_sys}. Most of the systematic uncertainties on the ST side cancel due to the tagging technique described in Section ~\ref{evt_sel}.

The uncertainty due to tracking efficiency is 1.0\% for each track, as determined from a study of the control samples of $J/\psi \rightarrow pK^{-}\bar{\Lambda}+c.c.$ and $J/\psi \rightarrow \Lambda \bar{\Lambda}$~\cite{2012SysTrkPPRD86}. The uncertainties arising from the differences of PID efficiencies between data and MC simulation for the proton (0.6\%) and pion (1.0\%) are determined with the control samples of $J/\psi \rightarrow \pi^{+}\pi^{-}p\bar{p}$ and $J/\psi \rightarrow \pi^{+}\pi^{-}\pi^{0}$~\cite{2019SysPIDPRD99}, respectively.
The uncertainty due to the $\Lambda$ vertex fit is determined to be 1\%~\cite{2012SysTrkPPRD86} by using the same control samples used in the tracking uncertainty estimation. The uncertainty associated with the $\pi^{0}$ reconstruction is derived from the control sample of $J/\psi \rightarrow \pi^{+}\pi^{-}\pi^{0}$ and assigned to be 2\%~\cite{2010SysPi0PRD81}.

In this analysis, the ST efficiency for reconstructing a tagged $\bar{\Xi}^{+}$ has been assumed to be independent of the multiplicities of the $\Xi^{-}$ side. To evaluate the potential bias of this assumption, we use simulated samples to study the tag efficiencies with two different decay modes of $\Xi^{-}$ ($\Xi^{-} \rightarrow \textrm{anything}$ and $\Xi^{-} \rightarrow \Xi^{0} e^{-} \bar{\nu}_e$), and take their difference (1.9\%) as the tag efficiency bias. The systematic uncertainties due to the mass window requirements for the $\Lambda$ and $\Xi^{0}$ and the momentum
of the $\Xi^{0}$ are 1.1\%, 4.2\%, and 1.2\%, respectively, as determined from the average impacts on the upper limit when varying these requirements from 1 to 3 standard deviations of their distributions. The systematic uncertainty due to the $q^{2}$ fitting range is estimated by varying the fitting range by 0.001 $({\rm GeV}/c^{2})^{2}$, and take its impact on the upper limit (1.6\%) as the systematic uncertainty. The uncertainties due to the cited branching fractions $\mathcal{B}(\Xi^{0}\rightarrow\Lambda\pi^{0})$, $\mathcal{B}(\Lambda \rightarrow p \pi^{-})$, and $\mathcal{B}(\pi^{0} \rightarrow \gamma\gamma)$
are 0.01\%, 0.78\%, and 0.03\%, respectively~\cite{2020PdgPTEP2020}. The systematic uncertainty due to the signal model is estimated by changing the model with a uniform distribution over the phase space to the one with the angular distribution~\cite{2019YangPRD92, 2003SigARNPS53}, and the effects from different signal models are negligible. The total systematic uncertainty is estimated to be 6.2\% by summing up all individual uncertainties in quadrature.

\begin{table}[htp]
\center
\caption{Relative systematic uncertainties for the branching fraction measurement.}
\label{tab:sum_sys}
\begin{tabular}{lc}
\hline\hline
Source                                                  & Uncertainty~(\%) \\ \hline
Tracking                                                & 2.0        \\
PID                                                     & 1.6        \\
$\Lambda$ vertex fit                                    & 1.0        \\
$\pi^{0}$ reconstruction                                & 2.0        \\
Tag efficiency bias                                     & 1.9        \\
Mass window of $\Lambda/\Xi^{0}$                        & 4.3        \\
Requirement on $p_{\Xi^{0}}$                            & 1.2        \\
Fitting range                                           & 1.6        \\
Cited branching fractions                               & 0.8        \\ \hline
Total                                                   & 6.2        \\ \hline \hline
\end{tabular}
\vspace{-0.2cm}
\end{table}

\section{\boldmath{RESULT}}
\begin{figure}[tb]

  \includegraphics[width=0.45\textwidth]{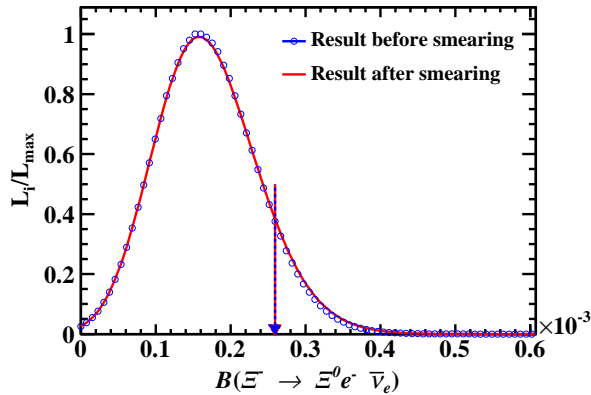}
  \caption{The likelihood ratio distribution as a function of $\mathcal{B}(\Xi^{-} \rightarrow \Xi^{0} e^{-} \bar{\nu}_{e})$ after (before) smearing with the systematic Gaussian function is the red solid (blue circle-shaped) curve. The red solid (blue dashed) arrow indicates the upper limit on the branching fraction at 90\% C.L. after (before) smearing with the systematic Gaussian function.}
\label{fig:UL}
\end{figure}

 No obvious signal is observed, and the upper limit on the DT yield for the $\Xi^{-} \rightarrow \Xi^{0} e^{-} \bar{\nu}_{e}$ decay is set at 90\% C.L. based on Eq.~\ref{tag_method} using a Bayesian method~\cite{2007Bayesian578}.
We perform a series of fits to the $q^{2}$ distribution by fixing the branching fraction of the signal process at different values, and scan the ratio of the resultant likelihood value (${\rm L_{i}}$) and the maximum likelihood value (${\rm L_{max}}$). To incorporate the effect of systematic uncertainties, the likelihood ratio distribution as a function of the branching fraction is then convolved with a Gaussian function, which has a width given by the overall systematic uncertainty, as shown in Fig.~\ref{fig:UL}.
The upper limit on the branching fraction at the 90\% C.L. is the value that yields 90\% of the likelihood ratio integral over the branching fraction from zero to infinity, and is determined to be
\begin{center}
$\mathcal{B}(\Xi^{-} \rightarrow \Xi^{0} e^{-} \bar{\nu}_{e})<2.59\times10^{-4}$.
\end{center}

\section{\boldmath SUMMARY}\label{summary}

In summary, based on a data sample with $(10.087\pm0.044)\times10^{9}$ $J/\psi$ events collected at $\sqrt{s} = 3.097$~${\rm GeV}/c^{2}$ with the BESIII detector, we search for the hyperon semileptonic decay $\Xi^{-} \rightarrow \Xi^{0} e^{-} \bar{\nu}_{e}$. No obvious signal is observed and the upper limit on the branching fraction is set to be $\mathcal{B}(\Xi^{-} \rightarrow \Xi^{0} e^{-} \bar{\nu}_{e})<2.59\times10^{-4}$ at 90\% C.L.
This result is one order of magnitude more strict than that of
BNL's measurement~\cite{1974YehPRD10}, and provides an important experimental constraint for the theoretical study of the SU(3) symmetry breaking mechanism.

\section*{\boldmath ACKNOWLEDGMENTS}

The BESIII collaboration thanks the staff of BEPCII and the IHEP computing center for their strong support. This work is supported in part by National Key R$\&$D Program of China under Contracts Nos. 2020YFA0406300, 2020YFA0406400; National Natural Science Foundation of China (NSFC) under Contracts Nos. 11625523, 11635010, 11735014, 11822506, 11835012, 11935015, 11935016, 11935018, 11961141012, 12022510, 12025502, 12035009, 12035013, 12061131003; the Chinese Academy of Sciences (CAS) Large-Scale Scientific Facility Program; Joint Large-Scale Scientific Facility Funds of the NSFC and CAS under Contracts Nos. U1732263, U1832207; CAS Key Research Program of Frontier Sciences under Contract No. QYZDJ-SSW-SLH040; 100 Talents Program of CAS; INPAC and Shanghai Key Laboratory for Particle Physics and Cosmology; ERC under Contract No. 758462; European Union Horizon 2020 research and innovation programme under Contract No. Marie Sklodowska-Curie grant agreement No 894790; German Research Foundation DFG under Contracts Nos. 443159800, Collaborative Research Center CRC 1044, GRK 214; Istituto Nazionale di Fisica Nucleare, Italy; Ministry of Development of Turkey under Contract No. DPT2006K-120470; National Science and Technology fund; Olle Engkvist Foundation under Contract No. 200-0605; STFC (United Kingdom); The Knut and Alice Wallenberg Foundation (Sweden) under Contract No. 2016.0157; The Royal Society, UK under Contracts Nos. DH140054, DH160214; The Swedish Research Council; U. S. Department of Energy under Contracts Nos. DE-FG02-05ER41374, DE-SC-0012069. This paper is also supported by the NSFC under Contract No. 11947038, U1932102 and the High Performance Computing Center of Henan Normal University.

\end{document}